# Status of the AMS Experiment

A. Kounine
*MIT, Cambridge, MA 02139, USA*

Alpha Magnetic Spectrometer (AMS-02) is a general purpose high energy particle detector which will be deployed on the International Space Station (ISS) at the end of 2010 – beginning of 2011 to conduct a unique 10 to 18 year mission of fundamental physics research in space. Among the physics objectives of AMS are a search for an understanding of Dark Matter, Antimatter, the origin of cosmic rays and the exploration of new physics phenomena not possible to study from ground based experiments. This article reviews the status of the AMS-02 detector construction, detector calibration results as well as AMS physics potential for new phenomena searches.

## 1. INTRODUCTION

Tremendous interest to space-borne particle physics experiments stems from the unique features of experimentation in space: possibility of studying primordial particles created in the early Universe in a clean, almost background free environment. Success of the first space-borne particle detectors such as IMP-5,7,8, HEAO-3, ACE, EGRET [1] lead to proposals of a more complex experiments (PAMELA, FERMI, AMS) [2,3]. These new experiments address the most intriguing questions of the modern cosmology – baryon asymmetry of the Universe and its mass density composition.

AMS is an international collaboration composed of 64 institutes from 16 countries. Construction of the detector is completed and it is expected to be launched by NASA to the International Space Station at the end of 2010. The technical goals of AMS are to reach a sensitivity of $\overline{He}/He = 1/10^{10}$, an $e^+/p$ rejection of $1/10^6$ and to measure the composition and spectra of charged particles with of an accuracy of 1%. This represents a considerable sensitivity improvement compared to the previous experiments with space-borne magnetic spectrometers.

There is a strong demand for precision measurements of cosmic rays in the energy region from 10 to 1000 GeV as the recent measurements of $e^+/(e^+ + e^-)$ by AMS-01, HEAT and PAMELA [4] indicate a large deviation of this ratio from the production of $e^+$ and $e^-$ predicted by a model that includes only ordinary cosmic ray collisions. These measurements are both at too low energy and of too limited statistics to shed the light on the origin of this significant excess. AMS is expected to provide definitive answers concerning the nature of this deviation.

## 2. AMS DETECTOR

AMS is a general purpose detector to study primordial cosmic ray particles in the energy range from 0.5 to 2000 GeV. In order to ensure that technologies used in the detector construction work reliably in space, a scaled down detector (AMS-01) was built, which was flown in 1998 on board the STS-91 mission for 10 days [3]. Final layout of the AMS-02 detector is presented in Figure 1. It consists of a Transition Radiation Detector; a hodoscope of Time of Flight counters; a Permanent Magnet; a precision silicon Tracker; an array of Veto Counters, surrounding Tracker; a Ring Image Cerenkov detector; and Electromagnetic Calorimeter.

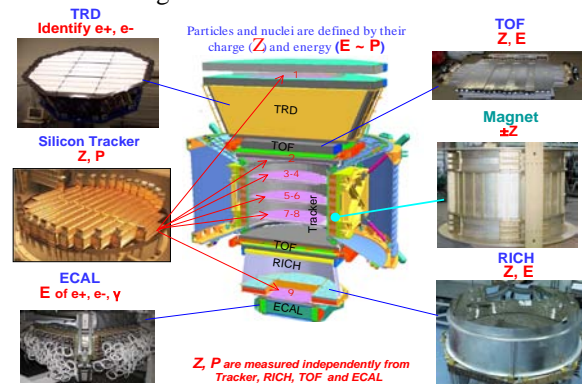

Figure 1: Layout of the AMS spectrometer

### 2.1. Transition Radiation Detector

The Transition Radiation Detector, TRD, is mounted on top of AMS (see Figure 1). It is designed to distinguish between light particles and heavy particles of equal charge and momentum, specifically between positrons and protons, which give the same signal in the silicon tracker but need to be separated to search for signals from dark matter annihilation in the cosmic ray spectra.

The TRD consists of 5248 straw tubes of 6 mm diameter. Sixteen straws are arranged in a module and mounted in 20 layers, each with a fiber fleece radiator of 20 mm thickness. Each straw tube is filled with a Xenon/$CO_2$ gas mixture; xenon efficiently captures the transition radiation generated in the radiator. Leak rate of the gas from TRD was measured to be 6µg/s. It is dominated by CO2 diffusion through the straw tube walls. With 5kg of CO2 onboard, this leak rate corresponds to the lifetime of over 24 years in space.

### 2.2. Time of Flight Counters

The Time of Flight, TOF, system of AMS-02 provides the fast trigger to the experiment, the measurement of the particle velocity including discrimination between upward and downward going particles and a measurement of the absolute charge.

Overall there are four planes of scintillators assembled into two mechanical structures - upper and lower TOF (see

Figure 1). The average time resolution of each counter has been measured to be 160 picoseconds, and the overall beta ($\beta$=v/c) resolution of the system has been measured to be 4% for $\beta\sim 1$ particles, according to the design specifications.

The Anti-Coincidence Counters (ACC) surround the AMS silicon tracker, just inside the inner cylinder of the vacuum case, to detect unwanted particles that enter or leave the tracker volume and induce signals close to the main particle track such that it could be incorrectly measured, for example confusing a nucleus trajectory with that of an anti-nucleus. The ACC consists of sixteen curved scintillator panels of 1 m length, instrumented with wavelength shifting fibers to collect the light and guide it to a connector from where a clear fiber cable guides it to the photomultiplier sensors mounted on the conical flange of the vacuum case.

### 2.3. Silicon Tracker and Permanent Magnet

The tracker is composed of 192 ladders, the basic unit that contains the silicon sensors, readout electronics and mechanical support. Three planes of honeycomb with carbon fiber skin, equipped with silicon ladders on both sides, constitute the inner part of the silicon tracker. Other three planes equipped with only one layer of silicon ladders are located on top of TRD, on top of the Permanent Magnet and in between Ring Image Cherenkov detector and Electromagnetic Calorimeters as indicated in Figure 1.

Each ladder has 100µm pitch silicon strips aligned with 3µm accuracy that measure coordinates of charged particles two orthogonal projections. Accuracy of the measurement in the bending plane is 10µm. Overall there are close to 200000 readout channels. Signal amplitude provides a measurement of the particle charge independent of other sub-detectors as presented in Figure 2.

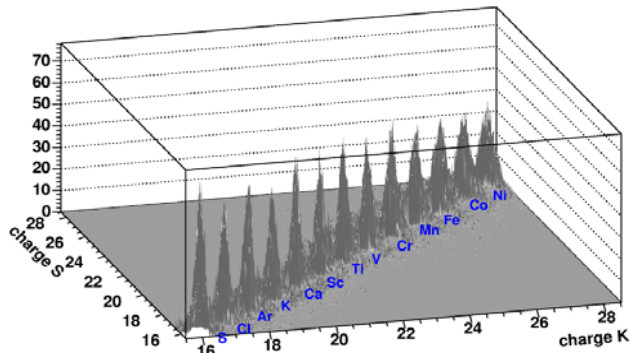

Figure 2: Correlation between bending plane amplitudes (charge S) and non-bending plane amplitudes (charge K) as measured in the heavy ion beam of 158 GeV/n.

Permanent Magnet with the central field of 1.4kG provides a bending power sufficient to measure protons up to Maximal Detectable Rigidity of 2.14TV. For He nuclei the Maximal Detectable rigidity is 3.75TV

### 2.4. Ring Imaging Cherenkov detector

The Ring Imaging Cerenkov (RICH) detector is designed to separate charged isotopes in cosmic rays by measuring velocities of charged particles with a precision of one part in a thousand. The detector consists of a dual dielectric radiator that induces the emission of a cone of light rays when traversed by charged particles with a velocity greater than that of the phase velocity of light in the material. The emitted photons are detected by an array of photon sensors after an expansion distance of 45 cm The measurement of the opening angle of the cone of radiation provides a direct measurement of the velocity of the incoming charged particle ($\beta$=v/c). By counting the number of emitted photons the charge (Z) of the particle can be determined (see Figure 3).

The radiator material of the detector consists of 92 tiles of silica aerogel (refractive index $n$=1.05) of 2.5 cm thickness and 16 tiles of sodium fluoride ($n$=1.33) of 0.5 cm thickness. This allows detection of particles with velocities greater than 0.953c and 0.75c respectively. The detection plane consists of 10,880 photon sensors with an effective spatial granularity of 8.5 x 8.5 mm$^2$. To reduce lateral losses the expansion volume is surrounded by a high reflectivity reflector with the shape of a truncated cone.

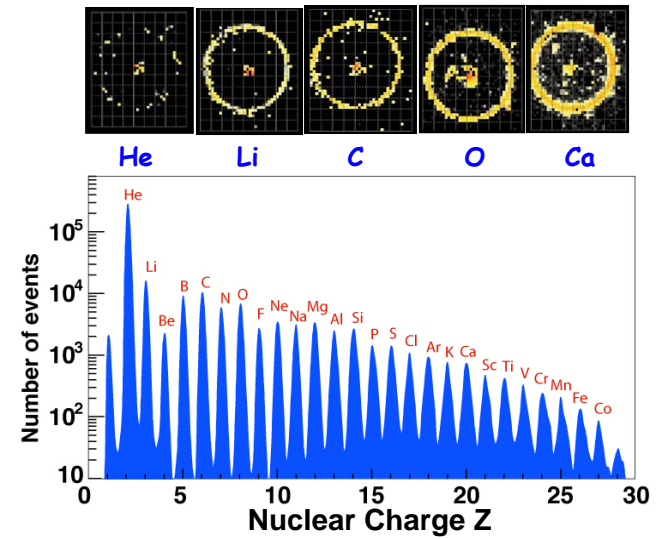

Figure 3: Shown on top are snapshots of the rings produced by the different nuclei as seen by RICH. Bottom figure is a spectrum of charges observed in 158 GeV/n heavy ion beam.

### 2.5. Electromagnetic Calorimeter

The AMS-02 electromagnetic calorimeter (ECAL) consists of a lead scintillating fiber sandwich with an active area of 648x648 mm$^2$ and a thickness of 166.5 mm. The calorimeter is composed of 9 superlayers, each 18.5 mm thick and made of 11 grooved, 1 mm thick lead foils interleaved with 10 layers of 1 mm diameter scintillating fibers. In each superlayer, the fibers run in one direction only. The 3-D imaging capability of the detector is obtained by stacking superlayers with fibers alternatively parallel to the X and Y axes (4 and 5 layers respectively). The calorimeter has a measured thickness corresponding to 17 radiation lengths.

Fibers are read out on one end by four anode Hamamatsu R7600-00-M4 photomultipliers (PMTs); each anode covers an active area of 9x9 mm$^2$, corresponding to 35 fibers, defined as a cell. In total the ECAL is subdivided into 1296 cells (324 PMTs) and this allows a sampling of the longitudinal shower profile by 18 independent measurements.

The signals are processed over a wide dynamic range, from one minimum ionizing particle, which produces about 8 photoelectrons per cell, up to the 60,000 photoelectrons produced in one cell by the electromagnetic shower of a 1 TeV electron. ECAL performance was studied in high energy electron and proton beams. The energy resolution for high energy electrons is measured to be 2-3% (see Figure 4), the angular resolution is ~1$^O$ and e/p separation is estimated to be ~10$^4$ for energies above 200 GeV

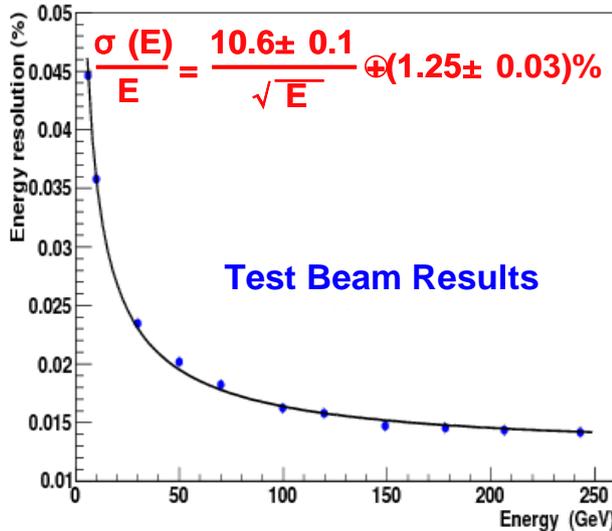

Figure 4: Energy resolution of ECAL as measured in stand-alone test with high energy electron beams.

## 3. CALIBRATION OF THE AMS DETECTOR

Various tests were performed with the individual AMS subsystems. This includes flight qualification tests: thermal, vibration, electromagnetic interference tests; as well as studies with different beams. In addition to that, several system wide tests were performed with the AMS detector: beam test at CERN and Thermal Vacuum test at ESTEC.

### 3.1. Beam test at CERN

From 4 to 9 February 2010, AMS was placed in the CERN test beam with 400 GeV protons and 180, 250 and 280 GeV electrons. AMS detector was installed on a support structure which allows 2 axes of translation and 2 axes of rotation for exposure of the detector to particles from all directions, as in space. Figure 5 summarizes the results from the data collected in the test beam by the integrated detector, showing that the track coordinate resolution is 10 µm (Figure 5a); the energy resolution for electrons is 2.5 to 3% (Figure 5b); the velocity resolution is 1/1000 with 400 GeV protons (Figure 5c); and that for 400 GeV protons TRD provides a proton rejection factor of 1/120 at 90% electron selection efficiency (Figure 5d). The combined proton rejection factor (TRD and ECAL) at 400 GeV was measured to be 10$^{-6}$.

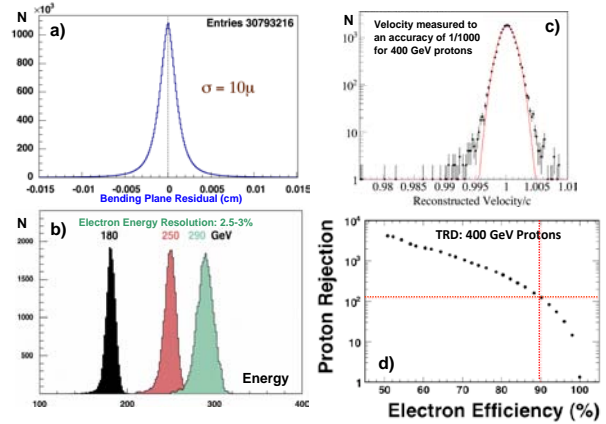

Figure 5: Summary of the beam test results: a) Tracker coordinate resolution; b) ECAL energy resolution; c) RICH velocity resolution; d) TRD e/p separation.

### 3.2. Thermal Vacuum test at ESTEC

AMS underwent Thermal Vacuum/Thermal Balance (TV/TB) testing in the Large Space Simulator at ESTEC, Holland in March – April 2010. The main objectives of the test were: testing performance of the sub-detectors under vacuum in a wide range of temperatures close to those on the Space Station; verification of the operability and the performance of the AMS Thermal Control Systems; and verification of the AMS Thermal Model. In particular, AMS heat rejection capability using radiators was verified. This is a very different mechanism compared to the convective cooling which occurs during the tests at ambient conditions.

Most components of AMS have already been tested in Thermal Vacuum chambers. The TV/TB test at the system level was required to verify the integrated performance of the detector. In particular, the functionality of AMS heaters and thermal interlocks was verified, including their impact on the overall AMS power consumption.

Heat load on the AMS cryostat in the Large Space Simulator was measured at two different vacuum case average temperatures: 242K and 260K, typical temperatures on the ISS. Stabilization of the AMS cryostat at an average vacuum case temperature of 242K is shown in Figure 6. Model predictions are indicated as a solid blue line for comparison. In general there is good agreement between the data and model predictions.

These measurements are used to estimate the endurance of the AMS cryostat on orbit. With the existing cryocoolers, purchased in 1999, the lifetime of AMS cryomagnet is estimated to be 20±4 months. The quoted uncertainty comprises both experimental and modeling errors as well as uncertainty related to the Space Station environmental conditions, *i.e.*, radiation from the nearby payloads, station altitude control and waiting time on the launch pad. Replacing the 11-year old cryocoolers with latest, same dimensions, model available from the same manufacturer (Sunpower, GT model), the expected endurance of the cryomagnet on ISS is 28±6 months. The cryocoolers are mounted on the Vacuum Case. Therefore they can be readily replaced without opening the magnet.

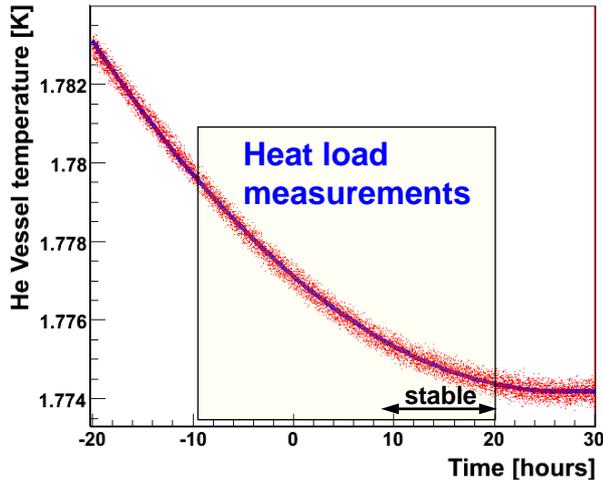

Figure 6: Temperature of the helium during stabilization at 242K (dots) and as modelled (line).

### 4. AMS MAGNET UPGRADE

The AMS experiment was constructed to operate on the International Space Station (ISS) for three years. The tests at the Thermal Vacuum Chamber of ESTEC show that the superconducting magnet with the new GT cryocoolers will have endurance of 28±6 months.

At the end of 2009 AMS was informed of discussions among ISS Space Agencies on the delay of de-orbiting the ISS from 2015 to 2020 and beyond. On March 11, 2010, there was an official announcement by NASA to continue the Space Station operation to at least 2020 and most likely 2028. AMS immediately activated plans to replace the superconducting magnet with the AMS-01 permanent magnet and to rearrange eight silicon tracker layers in order to keep the momentum resolution intact. A superconducting magnet was an ideal option for a three year stay on the ISS as originally planned for AMS. Termination of the Shuttle program and extension of the lifetime of the ISS to 2020-2028 eliminate any possibility of returning or refilling of AMS. Therefore, the superconducting magnet is no longer the ideal choice.

During the past ten years, the AMS-01 permanent magnet has been kept as a viable alternative to the superconducting magnet for AMS-02. After returning from the AMS-01 flight in June 1998, the permanent magnet was stored in a clean room and was shipped to RWTH Aachen in April 2010 for the installation into AMS-02 (Figure 7). The measurement of the field shows that in 12 years it has remained the same within 1%, corresponding to the accuracy of the measurements.

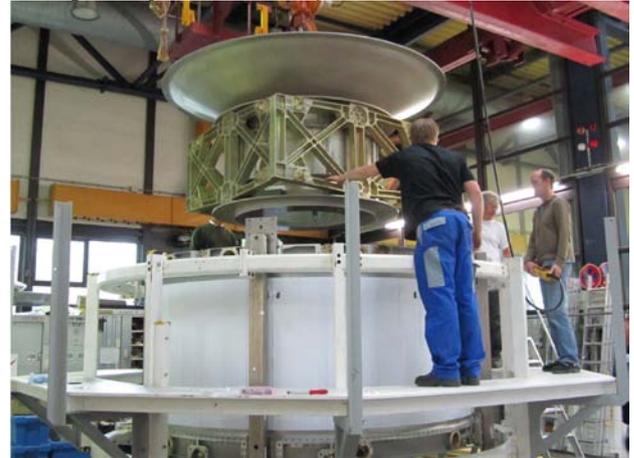

Figure 7: Installation of the Permanent Magnet into the AMS-02 support structure at RWTH, Aachen.

The advantage of using a permanent magnet is that it will be operational for the duration of the Space Station. The disadvantage is that it has a much smaller field – about 20% of that of the superconducting magnet. Therefore, the Tracker measuring arm was increased by moving on silicone plane on top of TRD and rearranging silicon ladders of other planes to make a new single-sided plane which was inserted between RICH and ECAL (see Figure 1). With this rearrangement the Maximal Detectable Rigidity of AMS is not affected (Figure 8).

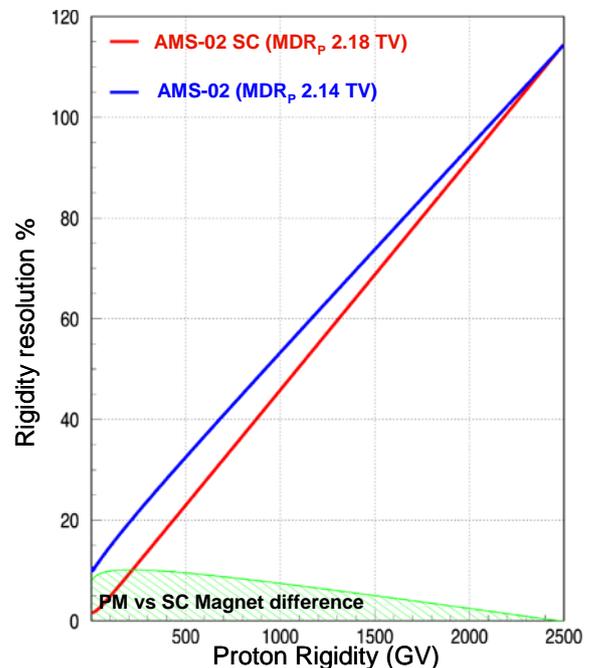

Figure 8: Rigidity resolution for the Superconducting Magnet (red line) and the Permanent magnet (blue line).

## 5. AMS PHYSICS POTENTIAL

AMS-02 is a general purpose particle detector capable of identifying and measuring simultaneously all cosmic ray particle species: photons, electrons, protons and nuclei as well as all corresponding anti-particles This feature becomes very important for distinguishing signals from new phenomena and background processes, given a significant uncertainty in the background calculations related to the modeling of the standard processes and subsequent propagation. AMS will measure spectra for nuclei in the energy range from 0.5 GeV/nucl to 2 TeV/nucl with 1% accuracy over the 11-year solar cycle. These spectra will constitute a stringent experimental test of the assumptions that go into the background estimates.

### 5.1. Dark Matter Searches

The most appealing candidate for Dark Matter is a stable neutralino which is a generic ingredient of SUSY models with a breaking scale of few hundred GeV. AMS-02 has potential to study neutralino annihilation using simultaneously four different final state particles: positrons, anti-protons, anti-deuterons and photons.

As seen on the insert in Figure 9, the available low energy measurements of the positron fraction indicate a strong deviation from the estimates based on the model that takes into account only cosmic ray collisions. AMS will measure all nuclei spectra thus providing stringent constraints on the background estimates.

Figure 9 shows the dependence of the measured positron spectrum on the neutralino mass. It is clearly seen that the structure in the spectrum due to the contribution of Dark Matter collisions is distinct up to neutralino masses of ~1 TeV. A large statistical sample from 10-18 years of data collection along with a good energy resolution of 2% implies that should this enhancement have a structure it can be clearly seen.

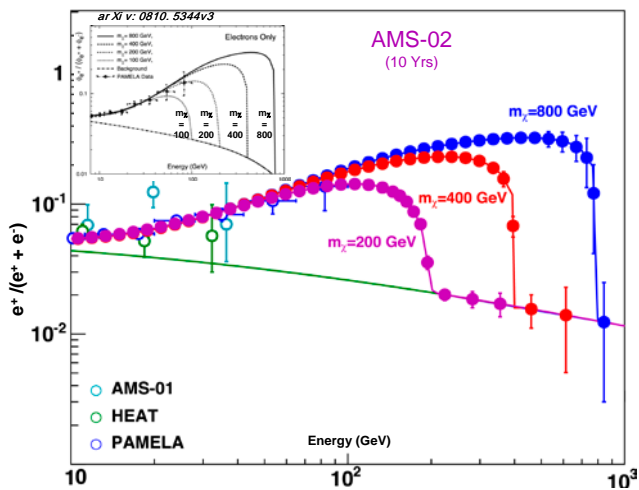

Figure 9: Dependence of the measured positron spectrum shape on the neutralino mass.

It should be noted that search for SUSY dark matter with AMS is complementary to the SUSY searches at LHC. Extensive analysis of the SUSY parameter space has revealed that some regions, not accessible to LHC [5], can be effectively explored with AMS [6]. Figure 10 shows the accuracy of AMS-02 measurements of $\bar{p}/p$ spectrum. Expected signal from Dark Matter collisions for $M\chi$ = 840 GeV (corresponding to benchmark M, not accessible to LHC) can be clearly identified as a structure on top of the background from ordinary cosmic rays.

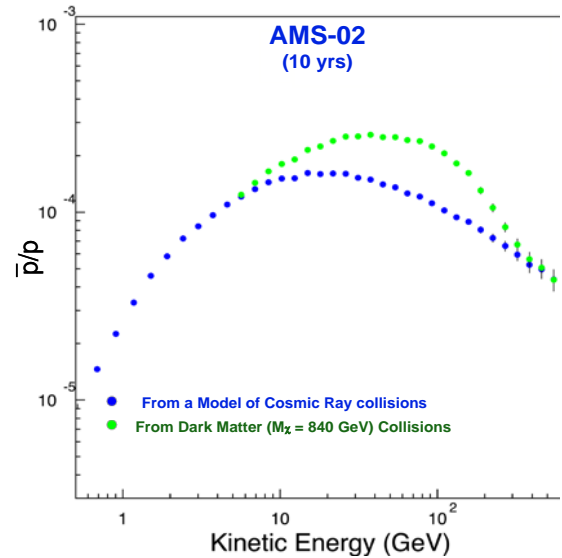

Figure 10: Accuracy of AMS-02 measurements of $\bar{p}/p$ spectrum for 10 years on ISS with the permanent magnet. Expected Dark Matter signal for $M\chi$ = 840 GeV (green dots with error bars) and background from ordinary cosmic ray collisions (blue dots with error bars).

Another possible Dark Matter candidate is Kaluza-Klein boson. Figure 11 shows a Dark Matter signal expected in the specific model of Eduardo Pontón and Lisa Randall [7]. This Figure highlights the sensitivity of AMS with the permanent magnet to detect 500 GeV Kaluza-Klein boson.

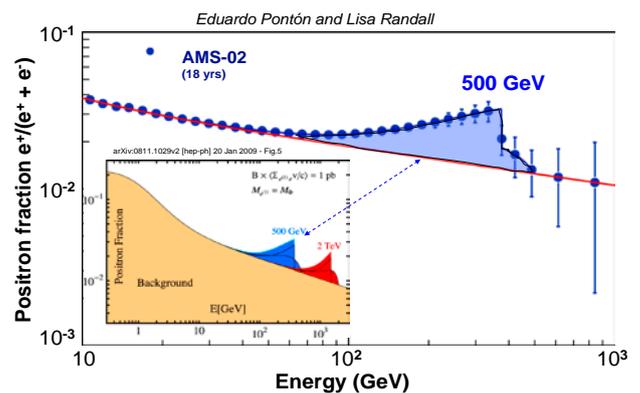

Figure 11: Expected spectrum from collisions of 500 GeV Kaluza-Klein bosons based on the model of Eduardo Pontón and Lisa Randall.

## 5.2. Anti-matter Searches

Conditions that are required for the baryon asymmetry in the Universe have no experimental support: to date neither strong CP-violation nor proton decays are observed experimentally. Therefore existence of large anti-matter domains in the Universe *a-priori* may not be ruled out. Such domains, if exist, would emit anti-particles which will eventually reach the Earth through a diffusion process. Production of anti-helium or heavier anti-nuclei in the interaction of ordinary matter in space is totally negligible; therefore observation of single anti-helium in space would constitute a strong argument in favor of such anti-matter domains.

One of the goals of AMS is to improve the sensitivity of a direct anti-matter search by 3 to 6 orders of magnitude and increase the current search range to 1 TeV as demonstrated in Figure 12. In this figure there are no assumptions that He and anti-He spectra are identical.

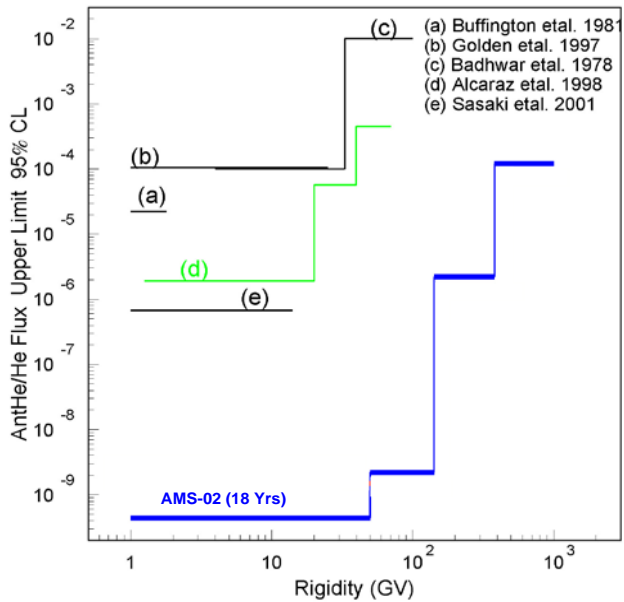

Figure 12: Sensitivity of AMS with the permanent magnet on ISS for 18 years compared with the other earlier measurements.

## 5.3. Strangelets

AMS represents an unprecedented opportunity to explore the unknown. One example is the search for new types of matter such as strangelets [8]. Strangelets, or strange quark matter (SQM), are new types of matter composed of three types of quarks (u, d, s) which may exist in the cosmos. Both lattice QCD and phenomenological bag model calculations indicate that SQM could be stable with lower energy levels than usual matter. SQM has a very low Z/A ratio, typically less than 0.13 compared to the ~0.5 of normal nuclei. Attempts to detect SQM production in accelerators are negative, which agrees with calculations that indicate they cannot be formed there by coalescence nor distillation (the minimum stable size, A>8, is too large). Neutron stars could in fact be one large strangelet at low "vapor" pressure, providing a source of SQM in cosmic rays. Searches for SQM on Earth and in lunar samples are negative but of limited sensitivity (e.g., large strangelets are so dense they would sink to the center of gravity). AMS-01 has observed a potential strangelet candidate with Z=2 and mass of 16.5 GeV, with the estimated flux of $5 \times 10^{-5}$ (sr $m^2$ s)$^{-1}$. AMS-02 will provide much improved sensitivity for the search of this new type of matter. Figure 13 shows the expected AMS-02 sensitivity for strangelet searches.

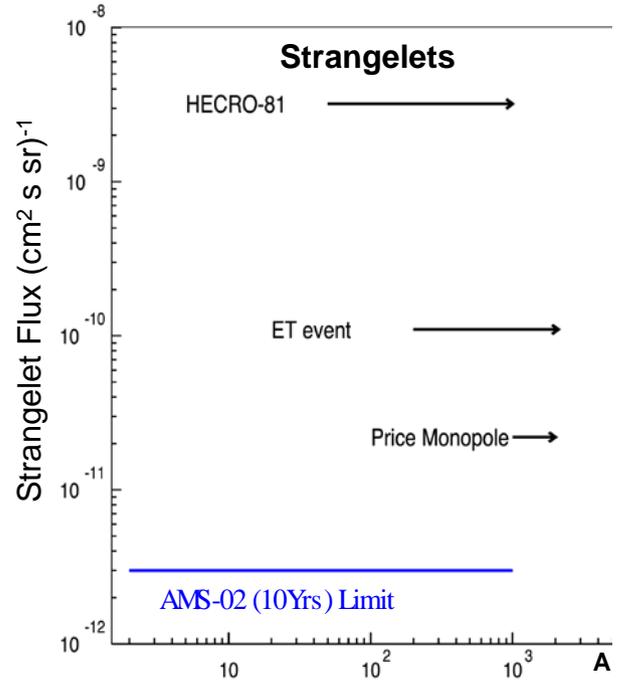

Figure 13. Sensitivity of a direct search for strangelets.

## References

[1] M.Garcia-Munoz, G.Mason and J.Simpson, *Ap.J.* **202**, 265 (1975); J.Engelmann *et al.*, *A&Ap.J.* **233**, 96 (1990); A.Labrador *et al.*, OG115, 28$^{th}$ ICRC, Tsukuba, Japan (2003); R.C.Hartman *et al.*, *Ap.JS.* **123**, 79 (1999).
[2] O.Adriani et al, AIP Conf. Proc. **1223** 33 (2010); B.Atwood *et al*, *ApJ* **697** 1071 (2009).
[3] M.Aguilar et al, Phys Rep **366/6** 331 (2002).
[4] S.W Berwick, *et al.*, Astrophys.J. **482** L191 (1997); M. Aguilar *et al*., Phys.Lett. B **646** 154 (2007); O.Adriani *et al.*, Nature **458** 607 (2009).
[5] M.Battaglia et al., preprint hep-ph/0306219; D.N.Spergel et al., preprint astro-ph/0603449.
[6] J.Ellis, private communication.
[7] E.Pontón and L.Randall, preprint *arXiv:0811.1029v2 (2009*)
[8] E.Witten, Phys.Rev. **D** 272 (1984).